# The Magnetic Origin of the Coherent Order Parameter in Hole-Doped Cuprates

A. Mourachkine

*Université Libre de Bruxelles, Service de Physique des Solides, CP233, Boulevard du Triomphe, B-1050 Brussels, Belgium*

___________________________________________________________________

**Abstract**

The phase diagram for hole-doped cuprates is discussed. By examining some recent inelastic neutron scattering data obtained on hole-doped cuprates we conclude that the coherent gap in hole-doped cuprates has most likely the magnetic origin and scales with $T_c$, on average, as $2\Delta_c/k_B T_c = 5.4$.

___________________________________________________________________

## 1. Introduction

The superconductivity requires the formation of the Cooper pairs and the phase coherence among them. In the BCS theory for conventional superconductors, the mechanisms responsible for the pairing and establishment of the phase coherence are identical: electrons couple to each other by phonons, and the phase coherence among the Cooper pairs is established also by phonons. Both phenomena occur almost simultaneously at $T_c$. In superconducting copper-oxides (high-$T_c$ superconductivity), there is a consensus that these two mechanisms occur at different temperatures, at least, in the underdoped regime, $T_{\text{pair}} \geq T_c$ [1-6]. The order parameters (OPs) responsible for each process have different dependencies on hole concentration, $p$ in $CuO_2$ planes [1,2]. The magnitude of the OP responsible for phase coherence, $\Delta_c$, which is proportional to $T_c$, has the parabolic dependence on $p$ [1,2,6,7]. While the magnitude of the OP responsible for pairing, $\Delta_p$, increases linearly with the decrease of hole concentration [1,2,8]. Both the $\Delta_c$ and $\Delta_p$ OPs are superconducting-like. However, there is no consensus on the origins of the two OPs. There is an evidence that the spin-exchange interactions play a central role in the cuprates [9,10]. Thus, it is reasonable to assume that one out of the two OPs has the magnetic origin.

Recent inelastic neutron scattering (INS) experiments have shown the presence of sharp magnetic collective mode ('resonance peak') in the superconducting state of $YBa_2Cu_3O_{6+x}$ (YBCO) [10-15]. The discovery of the resonance peak in $Bi_2Sr_2CaCu_2O_{8+x}$ (Bi2212) [16] points out that the resonance peak is an *intrinsic* feature of the superconductivity in the double-layer cuprates studied so far. The temperature dependence of the resonance peak demonstrates that this peak is intimately related to the establishment of superconductivity [17,10]. It is important to note that the $E_r$ is in quantitative agreement with the condensation energy of YBCO [18,19,10]. The resonance peak has been also observed by INS in a heavy fermion compound $UPd_2Al_3$ [20,21] for which spin fluctuations are believed to mediate the pairing interactions [22,23]. It is also important to note that the superconductivity in $UPd_2Al_3$ coexists with the antiferromagnetic order like in the cuprates. By making a parallel between the cuprates and heavy fermion compound $UPd_2Al_3$, the latter suggests that, in the hole-doped cuprates, there is an OP having the magnetic origin.

In the present work, we discuss the phase diagram and INS data obtained on hole-doped cuprates, and we conclude that the coherent OP, $\Delta_c$, has most likely the magnetic origin.

## 2. The coherent OP and INS data

Figure 1 shows a phase diagram for two energy gaps [24] in hole-doped cuprates [1,2]. In Fig. 1, the coherent gap, $\Delta_c$, scales with $T_c$ as $2\Delta_c/k_BT_c = 5.45$ [1,2]. The dependence $\Delta_c(p)$ is parabolic since $T_c = T_{c,\,max}[1 - 82.6(p - 0.16)^2]$, where $T_{c,\,max}$ is the maximum $T_c$ for each family of cuprates [7]. In Fig. 1, we present also INS data for YBCO [10-15] and Bi2212 [16], which can be found in Table I. One can see in Fig. 1 that there is a good agreement between the $\Delta_c$ and the INS data. So, it is reasonable to assume that the $\Delta_c$ has the magnetic origin.

In general, the superconductivity mediated by spin fluctuations implies that the coherent gap in hole-doped cuprates has the $d_{x^2-y^2}$ symmetry [9]. The average $2\Delta_c/k_BT_c$ value for the data presented in Table I is equal to 5.38.

## 3. Discussion

In $UPd_2Al_3$, the resonance-peak position $E_r = 0.36$ meV [20] does not match exactly the value of tunneling gap $2\Delta = 0.47$ meV [22] like it is in Fig. 1 for cuprates. There are, at least, two reasons for this. First of all, the INS measurements have been performed on $UPd_2Al_3$ single crystals whereas the tunneling measurements on thin films [20,22]. The second and, maybe, the most important reason for the small discrepancy between the tunneling and INS data is an anisotropic character of the energy gap which has most likely the d-wave symmetry [22,9]. The resonance peak in $UPd_2Al_3$ is detected at the antifferomagnetic Bragg point (0, 0, 0.5) [20], while the tunneling spectra are measured along the *c*-axis [22]. Thus, the INS and tunneling data obtained at different angles on the Fermi surface, and it is the main reason for the discrepancy of the data since the energy gap is anisotropic (d-wave).

The quasiparticles in conventional superconductors use phonons for pairing. Most likely, INS detects spin fluctuations which replace phonons [9]. Another explanation for the presence of the resonance peak in hole-doped cuprates is that it corresponds to the response of the magnetic domains in $CuO_2$ planes to the hole pairing [25]. It may be the case. However, the same response has been detected in the heavy fermion compound $UPd_2Al_3$ in which the superconductivity is mediated by spin fluctuations [22,23]. Thus, in both cases, this is an evidence for the presence of the magnetically-mediated superconductivity in hole-doped cuprates.

In order to explain the data, recently, we proposed a MCS (Magnetic Coupling of Stripes) model [6,2] which is based on the stripe model [26] which is in it's turn based on a spinon superconductivity along charge stripes. The main difference between the MCS and stripe models is that the coherent state of spinon superconductivity is established differently in the two models, by spin fluctuations into antiferromagnetic domains of $CuO_2$ planes in the MCS model, and by the Josephson coupling between stripes in the stripe model. Thus, in the MCS model, the superconductivity has two different mechanisms: along charge stripes for pairing and perpendicular to stripes for establishing the coherent state. As a consequence, carriers exhibit different properties in different directions: *fermionic* along charge stripes and *polaronic* perpendicular to stripes. We found that many experimental data can be explained by the MCS model [27]. Moreover, the MCS model may explain the s-wave superconductivity [28] in an electron-doped $Nd_{2-x}Ce_xCu_2O_4$ cuprate [6,27]. However, it is possible that there is another model of high-$T_c$ superconductivity which can explain experimental data better. Unfortunately, we are not aware of such model.

**4. Conclusions**

We discussed the phase diagram for the phase-coherence and pairing order parameters in hole-doped cuprates. By examining some recent inelastic neutron scattering data obtained on hole-doped cuprates we concluded that the coherent order parameter has most likely the magnetic origin and scales with $T_c$, on average, as $2\Delta_c/k_B T_c = 5.4$. We discussed also the MCS model of high-$T_c$ superconductivity in hole-doped cuprates.

## Acknowledgments

I thank R. Deltour for a discussion. This work is supported by PAI 4/10.

**Table I  INS data presented in Fig. 1**

| $YBCO_{6+x}$ | $T_c$ (K) | $E_r$ (meV) | $E_r / k_B T_c$ | Ref. |
|---|---|---|---|---|
| x = 0.5 | 52 | 25 | 5.6 | 11 |
| 0.6 | 62.7 | 34 | 6.3 | 12 |
| 0.7 | 67 | 33 | 5.72 | 11 |
| 0.8 | 82 | 39 | 5.53 | 10 |
| 0.83 | 85 | 35 | 4.8 | 13 |
| 0.92 | 91 | 41 | 5.24 | 13 |
| 0.97 | 92.5 | 40 | 5.03 | 14 |
| 1 | 92.4 | 41 | 5.16 | 15 |
| 1 | 93 | 41 | 5.13 | 11 |
| 1 | 89 | 39 | 5.1 | 13 |
| Bi2212 | 91 | 43 | 5.5 | 16 |

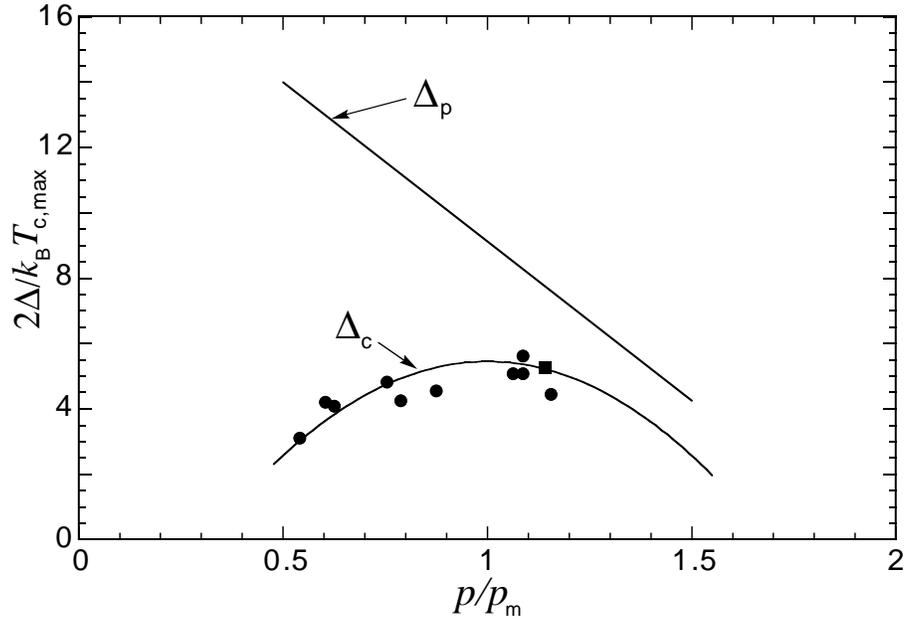

FIG. 1. Phase diagram for hole-doped cuprates: $\Delta_c$ is the coherent energy gap and $\Delta_p$ is the pairing energy gap [1]. INS data: dots (YBCO) and square (Bi2212). For detailed references, see Table I. The $p_m$ is a hole concentration with the maximum $T_c$.